\documentclass[12pt]{article}
\usepackage{amssymb}
\usepackage{amsmath}
\usepackage{bm}
\usepackage{amsthm}
\usepackage{pdfpages}
\usepackage{graphicx}
\usepackage{algorithmic, algorithm, enumerate}
\usepackage{float}

\newtheorem{theorem}{Theorem}

\input epsf.tex

\begin{document}

\begin{center}
{\Large \bf  R-optimal designs for multi-response regression models with multi-factors}

Pengqi Liu$^*$, Lucy Gao$^{**}$ and Julie Zhou$^{***}$\footnote{
Corresponding author, email: jzhou@uvic.ca,
phone: 250-721-7470.}

\vspace{5mm} 
* Department of Statistics and Data Science \\ 
Yale University, New Haven, CT, USA 06511 \\
\vspace{2mm} 
** Department of Biostatistics \\
University of Washington,  Seattle, WA, USA 98195-7232 \\ 
\vspace{2mm} 
*** Department of Mathematics and Statistics \\
University of Victoria, 
Victoria, BC, Canada V8W 2Y2 \\

\end{center}

\bigskip

\noindent
ABSTRACT

{\small 
We investigate R-optimal designs for multi-response regression models with multi-factors, where the
random errors in these models are correlated. Several theoretical results are derived for R-optimal designs, including scale invariance, reflection symmetry, line and plane symmetry, and dependence on the covariance matrix of the errors. All the results can be applied to linear and non-linear
models. In addition, an efficient algorithm based on an interior point method is developed for finding R-optimal designs on discrete design spaces. The algorithm is very flexible, and can be applied to any multi-response regression model. }



\newpage

\section{Introduction}



In multi-response experiment designs, multiple correlated responses are observed on each experimental unit. For example, we might observe multiple measures at a single time point from each experimental unit, or a single measure at multiple time points from each experimental unit. Multi-response experiment designs are commonly employed in applied sciences such as biomedical science, pharmaceutical science, chemistry, and engineering. 

Many papers have investigated optimal designs for the multi-response linear regression model with $m$ dependent response variables and $p$ independent design variables: 
\begin{eqnarray}
y_{ij}=\left( {\bf f}_i({\bf x}_j) \right)^\top {\bm \beta}_i+\epsilon_{ij}, ~i=1, \ldots, m, ~j=1, \ldots, n,
\label{model10}
\end{eqnarray}
where $y_{ij}$ is the $i$th response variable $y_i$ observed on the $j$th experimental unit, ${\bf x}_j$ is a $p$-vector denoting the value of the $p$ design variables on the $j$th experimental unit, ${\bf f}_i({\bf x})$ is a $q_i$-vector of known functions of ${\bf x}$,  $ {\bm \beta}_i \in R^{q_i}$ is the vector of unknown regression parameters for $y_i$, and $\epsilon_{ij}$ are random errors with mean zero. Vectors ${\bf f}_i({\bf x})$ and ${\bf f}_k({\bf x})$ may be different for $i \neq k$, and the total number of unknown regression parameters in (\ref{model10}) is $q=q_1+\cdots+q_m$.  We assume that observations on different experimental units are independent. Liu and Yue (2013) reviewed previous work on optimal designs for model \eqref{model10}. 



A-, c-, D- and E-optimal design criteria are commonly used in the optimal design literature, and the D-optimal design criterion is especially common. In contrast to the D-optimality criterion, which minimizes the volume of the confidence ellipsoid, the R-optimality criterion proposed by Dette (1997) minimizes the volume of the $q$-dimensional rectangle based on Bonferroni $t$-intervals. Practical applications typically use Bonferroni $t$-intervals, since Bonferroni $t$-intervals are easier to interpret and compute than confidence ellipsoids. Thus, R-optimal designs are an attractive alternative to D-optimal designs.

Many recent papers have investigated R-optimal designs for one-response models. 
Liu et al. (2014)
studied R-optimal designs for multi-factor models and showed that the product type designs are optimal for Kronecker product type models with complete interactions.
Liu et al. (2016)  
investigated R-optimal designs for
second-order response surface models, and an algorithm was proposed to find optimal designs 
on the $p-$dimensional unit cube or ball.
He and Yue (2018) 
investigated R-optimal designs for trigonometric regression models,
and He and Yue (2019) 
considered R-optimal designs for regression models with asymmetric errors using the second-order least squares estimator. Liu and Yue (2013) investigated R-optimal designs for model \eqref{model10} and obtained several results, including the convexity of loss function, the directional derivative of the loss function and the equivalence theorem.  Examples are given for two-response models, and R-optimal designs are constructed for some special cases of model \eqref{model10}, such as linear and quadratic response models. However, it is challenging to construct R-optimal designs for model \eqref{model10} unless ${\bf f}_i({\bf x})$ are simple functions. 


Many papers have proposed numerical algorithms for computing optimal designs. Atashgah and Seifi (2007, 2009) used semi-definite programming to construct D- and E-optimal designs for model \eqref{model10}. Wong et al. (2019) also used semi-definite programming to construct A-, A$_s$-, c- and D-optimal designs for multi-response linear and nonlinear regression models. However, these algorithms 
cannot be applied to construct R-optimal designs for model \eqref{model10}. 

In this paper, we investigate R-optimal designs for the multi-response linear regression model \eqref{model10} and locally R-optimal designs for multi-response non-linear regression models. We derive general theoretical results, including scale invariance, symmetry, and the relationship between
R-optimal designs and the error covariance matrix. Several of the theoretical results are new, and a few of the theoretical results are extensions of results in Liu and Yue (2013) and Dette (1997). We also propose a computationally efficient numerical algorithm based on an interior point method to compute R-optimal designs on discrete design spaces.  Our algorithm can be applied to any multi-response  linear/non-linear regression model.

The rest of the paper is organized as follows.
We describe the R-optimality criterion for model \eqref{model10} and investigate its theoretical properties in Section 2.
In Section 3, we develop an algorithm to compute R-optimal designs. 
Applications of R-optimal designs are in Section 4. Concluding remarks are in Section 5, and all proofs and derivations are in the Appendix.

\section{R-optimality and its properties}

To present model (\ref{model10}) in matrix form, we define the following vectors and matrices:
\begin{eqnarray*}
{\bf y}_j=\left( \begin{array}{c}
y_{1j} \\
y_{2j} \\
\vdots \\
y_{mj}
\end{array}
\right)_{m \times 1}, ~~
{\bm \epsilon}_j=\left( \begin{array}{c}
\epsilon_{1j} \\
\epsilon_{2j} \\
\vdots \\
\epsilon_{mj}
\end{array}
\right)_{m \times 1}, ~~
{\bf Y}=\left( \begin{array}{c}
{\bf y}_{1} \\
{\bf y}_{2} \\
\vdots \\
{\bf y}_{n}
\end{array}
\right)_{mn \times 1}, ~~
{\bm \epsilon}=\left( \begin{array}{c}
{\bm \epsilon}_{1} \\
{\bm \epsilon}_{2} \\
\vdots \\
{\bm \epsilon}_{n}
\end{array}
\right)_{mn \times 1}, 
\end{eqnarray*}

\begin{eqnarray*}
Z({\bf x}_j)=\left(
\begin{array}{cccc}
{\bf f}_1^\top({\bf x}_j) &  0 & \cdots & 0 \\
0 & {\bf f}_2^\top({\bf x}_j) &  \cdots & 0 \\
\vdots  & \vdots & \ddots & \vdots  \\
0 & 0 & \cdots & {\bf f}_m^\top({\bf x}_j)
\end{array}
\right)_{m \times q}, ~~
{\bf Z}=\left( \begin{array}{c}
 Z({\bf x}_1) \\
 Z({\bf x}_2) \\
\vdots \\
 Z({\bf x}_n)
\end{array}
\right)_{mn \times q}, ~~
{\bm \beta}=\left( \begin{array}{c}
{\bm \beta}_{1} \\
{\bm \beta}_{2} \\
\vdots \\
{\bm \beta}_{m}
\end{array}
\right)_{q \times 1}. 
\end{eqnarray*}
These definitions are similar to those in Wong et al. (2019). Then, model (\ref{model10}) can be written as ${\bf Y}={\bf Z} {\bm \beta} + {\bm \epsilon},$
where the covariance matrix of the errors ${\bm \epsilon}$ is given by
\begin{eqnarray*}
{\bf V}=\mbox{Cov}({\bm \epsilon}) ={\bf V}_0 \oplus {\bf V}_0  \oplus \cdots \oplus {\bf V}_0~~~
\mbox{with}~~
{\bf V}_0=\mbox{Cov}({\bm \epsilon}_1).
\end{eqnarray*}
The notation $\oplus$ denotes the matrix direct sum. 
The best linear unbiased estimator (BLUE) of ${\bm \beta}$
is the generalized least squares estimator (GLSE) given by 
\begin{eqnarray} 
\hat{\bm \beta} = \left({\bf Z}^\top {\bf V}^{-1}{\bf Z} \right)^{-1}  {\bf Z}^\top {\bf V}^{-1}{\bf Y}, \label{glse}
\end{eqnarray}
and its covariance matrix is given by 
\begin{eqnarray}
\mbox{Cov}(\hat{\bm \beta}) =\left({\bf Z}^\top {\bf V}^{-1}{\bf Z} \right)^{-1} 
=\left( \sum_{j=1}^n \left(Z({\bf x}_j)\right)^\top {\bf V}_0^{-1} Z({\bf x}_j) \right)^{-1}.
\label{Cove2}
\end{eqnarray}

A design measure on a discrete design space $S_N = \{ {\bf u}_1, {\bf u}_2, \ldots, {\bf u}_N\} \subset R^p$ is written as
\begin{eqnarray*}
\xi({\bf w}) = \left( \begin{array}{cccc}
{\bf u}_1 & {\bf u}_2 & \ldots & {\bf u}_N \\
w_1 & w_2 & \ldots & w_N
\end{array}
\right),
\end{eqnarray*}
where weight vector ${\bf w}=(w_1, w_2, \ldots, w_N)^\top$ satisfies $w_j \ge 0$ and $\sum_{j=1}^N w_j=1$.
A point ${\bf u}_j$ is called a support point of $\xi({\bf w})$ if $w_j > 0$.
Design points ${\bf x}_1, \ldots, {\bf x}_n$ in (\ref{model10}) are selected from distribution $\xi({\bf w})$ by
rounding $nw_j$ into integers, say $[nw_j]$, such that $\sum_{j=1}^N  [nw_j] =n$. Let
\begin{eqnarray}
{\bf I}({\bf w})=\sum_{j=1}^N w_j U_j^\top {\bf V}_0^{-1} U_j ~~~\mbox{and}~~~ {\bf A}({\bf w})= {\bf I}^{-1}({\bf w}), 
\label{matrixA}
\end{eqnarray}
where $U_j=Z({\bf u}_j)$ for $j=1, \ldots, N$. From \eqref{Cove2} and \eqref{matrixA},  $\mbox{Cov}(\hat{\bm \beta})$ is proportional to ${\bf A}({\bf w})$.


We focus on approximate designs on $S_N$.
Let ${\bf e}_r$ be the $r$th unit vector in $R^q$, $r=1, \ldots, q$. An approximate R-optimal design for model \eqref{model10} on $S_N$ minimizes 
the following loss function:
$$\Phi_R({\bf w}) = \prod_{r=1}^q {\bf e}_r^\top {\bf A}({\bf w}) {\bf e}_r.$$
Equivalently, we can minimize the following logarithmic loss function:
\begin{eqnarray}
\phi({\bf w})=\mbox{log} \left( \Phi_R({\bf w})\right) = \sum_{r=1}^q \mbox{log} \left({\bf e}_r^\top {\bf A}({\bf w}) {\bf e}_r \right),
\label{logloss}
\end{eqnarray}
where $\log(\cdot)$ denotes the natural logarithm function. 
Loss functions $\Phi_R({\bf w})$ and $\phi({\bf w})$ are defined to be infinite if ${\bf I}({\bf w})$ is singular. Liu and Yue (2013) focused on a loss function similar to $\Phi_R({\bf w})$ to obtain the equivalence theorem and other results. For one-response 
models, theoretical results were also obtained using  a loss function similar to $\Phi_R({\bf w})$. In this paper, we focus on $\phi({\bf w})$ instead of $\Phi_R({\bf w})$, which facilitates the derivation of new theoretical results in Section \ref{r-properties}. Another advantage of focusing on $\phi({\bf w})$ is that the derivative of $\phi({\bf w})$ can be computed more efficiently than the derivative of $\Phi_R({\bf w})$. This will be helpful for the numerical algorithms in Section 3.

\subsection{Convexity and equivalence result for R-optimal design}

We study the convexity of $\phi({\bf w})$ and obtain the equivalence result for R-optimal designs for model \eqref{model10}. For any two weight vectors ${\bf w}_1$ and ${\bf w}_2$ in $R^N$ and  $\alpha \in [0, 1]$, we denote the convex combination of ${\bf w}_1$ and ${\bf w}_2$ as ${\bf w}_{\alpha}=(1-\alpha) {\bf w}_1 + \alpha {\bf w}_2.$	
\begin{theorem}
For any two weight vectors ${\bf w}_1$ and ${\bf w}_2$ in $R^N$ that ${\bf I}({\bf w}_1)$ and ${\bf I}({\bf w}_2)$ are non-singular,
$\phi({\bf w}_\alpha)$ is a convex function of $\alpha$, where $\alpha \in [0, 1]$.
\label{Th1}
\end{theorem}

The proof of Theorem \ref{Th1} is in the Appendix. Since $\phi({\bf w}_\alpha) = \log (\Phi_R({\bf w}_\alpha))$, by Theorem \ref{Th1}, $\Phi_R({\bf w}_\alpha)$ is a log-convex function of $\alpha$. Log-convex functions are convex, but convex functions are not necessarily log-convex (Boyd and Vandenberghe 2004, p104). Thus, 
Theorem \ref{Th1} establishes a stronger property of $\Phi_R({\bf w_\alpha})$ than convexity of $\Phi_R({\bf w_\alpha})$. 

We  derive the equivalence result for R-optimal designs on $S_N$ using Theorem 1.
For $j=1, \ldots, N$ and any weight vector ${\bf w}$ that ${\bf I}({\bf w})$ is non-singular, let
\begin{eqnarray}
d({\bf w},j)=\mbox{trace} \left( {\bf A}({\bf w}) U_j^\top  {\bf V}_0^{-1} U_j {\bf A}({\bf w}) 
\sum_{r=1}^q \frac{{\bf e}_r {\bf e}_r^\top}{{\bf e}_r^\top {\bf A}({\bf w}){\bf e}_r} \right) - q.
\label{eqv1}
\end{eqnarray}

\begin{theorem}
Design $\xi({\bf w}^*)$ is R-optimal for model (\ref{model10}) on $S_N$ if and only if
$d({\bf w}^*,j) \le 0$
 for all $j=1, \ldots, N$, and the equality holds at the support points of $\xi({\bf w}^*)$.
\label{Th2}
\end{theorem}

The proof of Theorem \ref{Th2} is similar to the derivation in Kiefer (1974) 
and is given in the Appendix. This result is also consistent with that of Liu and Yue (2013)
and can be used to verify the R-optimality of a design. In Section 3,
for a small positive $\delta$, we will relax the condition in Theorem \ref{Th2} for designs computed numerically to 
\begin{eqnarray}
d({\bf w}^*,j) \le \delta, ~~\mbox{for~} j=1, \ldots, N.
\label{eqv2}
\end{eqnarray}

\subsection{Properties of R-optimal designs}
\label{r-properties}
We explore several theoretical properties of R-optimal designs for model \eqref{model10} including scale invariance, symmetry, and dependence on the error covariance matrix ${\bf V}_0$.

Consider a scale transformation $T$ mapping the design space $S_N$ to another space, denoted by 
$S_N^T=\left\{ T{\bf u}_1, T{\bf u}_2, \cdots, T{\bf u}_N \right\}$,
where $T$ is a diagonal matrix with positive diagonal elements $t_1, \ldots, t_p$, so that $T$ scales the $p$ design variables by factors $t_1, \ldots, t_p$, respectively.
Note that $S_N^T$ still has $N$ distinct points.
If $\xi({\bf w}^*)$ with ${\bf w}^*=(w_1^*, \ldots, w_N^*)^\top$ 
is an R-optimal design on $S_N$ and 
the following design,
\begin{eqnarray*}
\left( \begin{array}{cccc}
T{\bf u}_1 & T{\bf u}_2 & \ldots & T{\bf u}_N \\
w_1^* & w_2^* & \ldots & w_N^*
\end{array}
\right)
,\end{eqnarray*}
is an R-optimal design on $S_N^T$, then the R-optimal design is \emph{scale invariant} under 
transformation $T$. The following result provides a sufficient condition on $f_i({\bf x}_j)$ for the scale invariance property of R-optimal designs for model \eqref{model10}. 

\begin{theorem}
Consider model (\ref{model10}) and a scale transformation $T$.  If there exists a non-singular diagonal matrix $Q$ 
such that $Z(T{\bf x})=Z({\bf x}) ~Q$ for all ${\bf x} \in S_N$  and $Q$ does not depend on ${\bf x}$,
then the R-optimal design is scale invariant under transformation $T$.
\label{Th3}
\end{theorem}

The proof of Theorem \ref{Th3} is in the Appendix. When the condition in Theorem \ref{Th3} can be verified, then it is easy to compute the R-optimal design on any scaled design space $S_N^T$ using the R-optimal design on $S_N$. This greatly reduces the computational burden of computing optimal designs for various sizes of design spaces.

We now study reflection symmetry of R-optimal designs.
Let $T_l$ be a reflection transformation for ${\bf x}$ with respect to variable $x_l$, for some $l \in \{1, 2, \ldots, p\}$.
This means that $T_l {\bf x} =(x_1, \ldots, x_{l-1}, -x_l, x_{l+1}, \ldots, x_p)^\top$.
Denote the transformed design space under $T_l$ by $S_N^{T_l}=\{ T_l {\bf u}_1,  T_l {\bf u}_2, \ldots,  T_l {\bf u}_N\}$.
If $S_N^{T_l}=S_N$, then  $S_N$ has a reflection symmetry with respect to variable $x_l$. 
For a $S_N$ that has a reflection symmetry with respect to variable $x_l$,
if $\xi({\bf w}^*)$ is an R-optimal design on $S_N$ 
 and it is the same as 
 the following design, 
\begin{eqnarray*}
\tilde{\xi}({\bf w}^*)=	\left( \begin{array}{cccc}
		T_l{\bf u}_1 & T_l{\bf u}_2 & \ldots & T_l{\bf u}_N \\
		w_1^* & w_2^* & \ldots & w_N^*
	\end{array}
	\right), 
\end{eqnarray*}
then $\xi({\bf w}^*)$ has a 
reflection symmetry with respect to variable $x_l$. Reflection symmetry is a useful property; if an R-optimal design with reflection symmetry with respect to $k$ variables, then the number of unknown weights in ${\bf w}$ can be reduced to about $N/(2^k)$. This can significantly reduce computational time. The following result provides a sufficient condition for reflection symmetry of R-optimal designs for model \eqref{model10}. 

\begin{theorem}
	For model (\ref{model10}) with a $S_N$ that has a reflection symmetry with respect to variable $x_l$, if there exists a diagonal matrix $Q$ 
	such that $Z(T_l{\bf x})=Z({\bf x}) ~Q$ for all ${\bf x} \in S_N$  and the diagonal elements of $Q$ are either  $+1$ or $-1$,
	then there exists an R-optimal design on $S_N$ that has a 
	reflection symmetry with respect to variable $x_l$.
	\label{Th4}
\end{theorem}

The proof of Theorem \ref{Th4} is in the Appendix.  Theorem \ref{Th4} can be applied   multiple times if $S_N$  has reflection symmetry with respect to several variables. 

The following example is used to illustrate 
Theorems \ref{Th3} and \ref{Th4}.

\noindent{\bf Example 1}  Consider a 3-response model with two design variables $x_1$ and $x_2$,
\begin{eqnarray*}
&&{\bf f}_1({\bf x})=(1, x_1, x_2, x_1x_2, x_1^2, x_2^2)^\top, \\
&&{\bf f}_2({\bf x})=(1, x_1, x_2, x_1x_2, x_1^2, x_2^2 )^\top, \\
&&{\bf f}_3({\bf x})=(1, x_1, x_2)^\top, 
\end{eqnarray*}	
and two design spaces
\begin{eqnarray*}
&&S_{N,1}=\left\{ (x_1,x_2) ~|~ x_1 \in S_{N_1}(0,1), ~
 x_2 \in S_{N_2}(0,1), ~N=N_1N_2  \right\}, \\
&&	S_{N,2}=\left\{ (x_1,x_2) ~|~ x_1 \in S_{N_1}(-1,1), ~
x_2 \in S_{N_2}(-5,5), ~N=N_1N_2    \right\},
\end{eqnarray*}	
where $S_N(a,b)=\left\{ a, a+\frac{b-a}{N-1}, a+\frac{2(b-a)}{N-1}, \ldots, a+\frac{(N-2)*(b-a)}{N-1}, b \right\}$
 denotes the set with $N$ equally spaced grid points in $[a, b]$,
 and $N_1$ and $N_2$ are two positive integers.
Let the scale transformation be $T{\bf x}=(t_1x_1, t_2x_2)^\top$ for positive $t_1$ and $t_2$.
It is easy to show that the condition in Theorem \ref{Th3} holds with matrix
$Q$ having diagonal elements: $1, t_1, t_2, t_1t_2, t_1^2, t_2^2, 
1, t_1, t_2, t_1t_2, t_1^2, t_2^2, 1, t_1, t_2$. Thus,
R-optimal designs for this model are scale invariant.
Design space $S_{N,2}$ has the reflection symmetry with respect to both variables $x_1$ and $x_2$, but 
$S_{N,1}$ does not. For the reflection symmetry with respect to $x_1$,
it is also easy to verify that the condition in Theorem \ref{Th4} holds with matrix
$Q$ having diagonal elements: $1, -1, 1, -1, 1, 1, 1, -1, 1, -1, 1, 1, 1, -1, 1$.
Thus, there exists an R-optimal design on $S_{N,2}$ that has the 
reflection symmetry with respect to $x_1$.  Similarly, 
there exists an R-optimal design on $S_{N,2}$ that has the 
reflection symmetry with respect to $x_2$ .  Combining the two results, we know that there exists an
R-optimal design having the 
reflection symmetry with respect to both $x_1$ and $x_2$.
Suppose a point $(a_0, b_0) \in S_{N,2}$, then $(-a_0, b_0), (a_0, -b_0)$ and $(-a_0, -b_0)$ are its reflection symmetric points with respect to $x_1$ and $x_2$. Then  these points,
$(a_0, b_0), (-a_0, b_0), (a_0, -b_0)$ and $ (-a_0, -b_0)$, have the same weight in the
R-optimal design having the 
reflection symmetry with respect to both $x_1$ and $x_2$.
R-optimal designs for this model with various $N$ and ${\bf V}_0$ are presented in Section 4 after we discuss numerical methods.
\hfill{$\Box$}

Other types of symmetry such as rotational symmetry, line symmetry and plane symmetry can also be explored for R-optimal designs. For some special cases of model (\ref{model10}) and design spaces $S_N$, R-optimal designs have the aforementioned symmetry properties. The derivations are similar to the proof of Theorem \ref{Th4}.  For instance, in Example 1 the design space $S_{N,1}$ with $N_1=N_2$ has a line symmetry with respect to line $x_1-x_2=0$. It can be easily shown that there exists an R-optimal design having a line symmetry.  

Furthermore, for a given symmetry of interest, let $T{\bf x}$ be the symmetric point of ${\bf x}$, where $T$ is a $p \times p$
corresponding matrix.  If $S_N^T=\{ T{\bf x} ~|~  {\bf x} \in S_N \}$ is the same as $S_N$, then
we say that the design space has the specified symmetry.
Under this symmetry, modify the  loss function in (\ref{logloss}) based on all the points in $S_N^T$ to get 
$$\tilde{\phi}({\bf w}) = \sum_{r=1}^q \log \left( {\bf e}_r^\top 
	\left( 
	\sum_{j=1}^N w_j  (Z(T{\bf u}_j))^\top {\bf V}_0^{-1} Z(T{\bf u}_j)  \right)^{-1}  
	 {\bf e}_r \right),    $$
where all ${\bf u}_j \in S_N$. 
If ${\phi}({\bf w}) = \tilde{\phi}({\bf w})$ for all ${\bf w}$, then there exists an R-optimal design
having the symmetry property on $S_N$. The proof is similar to the proof of Theorem \ref{Th4}.

The results in Theorems \ref{Th3} and \ref{Th4} and other symmetry properties do not depend on the error covariance matrix ${\bf V}_0$.  We can further explore the relationship between R-optimal designs and ${\bf V}_0$; three new results are derived for R-optimal designs below.

\begin{theorem}
	If the $m$ response functions in model (\ref{model10}) are the same, i.e., 
	${\bf f}_1({\bf x}) = \ldots ={\bf f}_m({\bf x})$ for all ${\bf x} \in S_N$,
	then the R-optimal design for  model (\ref{model10})  
	does not depend on the error covariance matrix ${\bf V}_0$.  In addition, 
	the R-optimal design for the multi-response is the same as that for a single 
	response model with the response function $\left({\bf f}_1({\bf x})\right)^\top {\bm \beta}_1$ based on the
	least squares estimator.	
	\label{Th5}
\end{theorem}

The proof of Theorem \ref{Th5} is similar to the proof in Wong et al. (2019, Theorem 3) and is omitted.
When ${\bf f}_1({\bf x}), \ldots, {\bf f}_m({\bf x})$ are different, R-optimal designs based
on multi-response models are usually different from 
those based on one-response models, and they often depend on ${\bf V}_0$. 
When an R-optimal design depends on ${\bf V}_0$,  it turns out that it depends on ${\bf V}_0$ only through its correlation matrix ${\bf R}_0 =  {\bm \Sigma}_0^{-1} {\bf V}_0 {\bm \Sigma}_0^{-1}$, 
where ${\bm \Sigma}_0$ is an $m \times m$ diagonal matrix with diagonal elements $\sigma_1, \ldots, \sigma_m$,
and $\sigma_1^2, \ldots, \sigma_m^2$ are the diagonal elements of ${\bf V}_0$.

\bigskip

\begin{theorem} 
 If an R-optimal design for (\ref{model10})  depends on ${\bf V}_0$, then it only depends on 
 the correlation matrix ${\bf R}_0$.
	\label{Th6}
\end{theorem}

The proof of Theorem \ref{Th6} is in the Appendix. 
One implication of Theorem \ref{Th6} is that all the properties of R-optimal designs discussed in Sections 2.1 and 2.2 hold if we change ${\bf V}_0$  to ${\bf R}_0$  in (\ref{matrixA}) 
and in the loss function $\phi({\bf w})$ in (\ref{logloss}). Furthermore, Theorem \ref{Th6} is useful when studying the sensitivity of R-optimal designs to ${\bf V}_0$. This is because ${\bf R}_0$ can be parameterized by fewer parameters than ${\bf V}_0$. For example, when $m=2$, 
${\bf R}_0=\left( \begin{array}{cc}
1 & \rho \\
\rho & 1
\end{array} 
\right),$
and we only need to study the sensitivity of R-optimal designs to the value of $\rho$. In fact, it turns out that R-optimal designs only depend on $\rho$ through $|\rho|$ for $m=2$, which is a consequence of the following general result about the dependence on ${\bf R}_0$.

\begin{theorem} 
Suppose that ${\bf R}_0$ and ${\bf R}_1$ are two possible $m \times m$ error correlation matrices for 
(\ref{model10}).  If there exists an $m \times m$ diagonal matrix ${\bf Q}_1$ with diagonal elements 
being either 1 or $-1$ such that
${\bf R}_0 =   {\bf Q}_1 {\bf R}_1 {\bf Q}_1$,
then the R-optimal design with correlation matrix ${\bf R}_0$  is the same as that with 
correlation matrix ${\bf R}_1$.
	\label{Th7}
\end{theorem}

The proof of Theorem \ref{Th7} is similar to that of Theorem \ref{Th6}.
For $m=2$,
it is clear that 
$${\bf R}_0=\left( \begin{array}{cc}
1 & \rho \\
\rho & 1
\end{array} 
\right) = \left( \begin{array}{cc}
1 & 0 \\
0 & -1
\end{array} 
\right)
{\bf R}_1
\left( \begin{array}{cc}
1 & 0 \\
0 & -1
\end{array} 
\right), ~~\mbox{with}~~
{\bf R}_1=\left( \begin{array}{cc}
1 & -\rho \\
-\rho & 1
\end{array} 
\right).$$
Thus, by Theorem \ref{Th7}, the R-optimal design for $m=2$ only depends on $|\rho|$. This special case
is discussed in one example in Liu and Yue (2013).

For $m>2$, Theorem \ref{Th7} can be applied to explore various correlation matrices and obtain interesting results. For instance, when $m=3$, consider
$${\bf R}_0=\left( \begin{array}{ccc}
1 & \rho_1 & 0 \\
\rho_1 & 1 & \rho_2 \\
0 & \rho_2 & 1
\end{array} 
\right), ~~\mbox{and}~~
{\bf R}_1=\left( \begin{array}{ccc}
1 & -\rho_1 & 0 \\
-\rho_1 & 1 & -\rho_2 \\
0 & -\rho_2 & 1
\end{array} 
\right).$$ 
Let ${\bf Q}_1$ be a diagonal matrix with diagonal elements $1, -1, 1$. Then we have 
${\bf R}_0 =   {\bf Q}_1 {\bf R}_1 {\bf Q}_1$.  Thus, R-optimal designs with
${\bf R}_0$ and ${\bf R}_1$ are the same, and they only depend on the absolute values of the correlations.
This result can be easily extended for $m>3$ and various  ${\bf R}_0$.

\section{Numerical computation of R-optimal designs}

 R-optimal designs have been derived for simple regression models with one response variable; see, for example,  
Dette (1997),
Liu et al. (2014),
and He and Yue (2018).
However, it is difficult to analytically construct optimal designs for 
complicated models. For A-, A$_s$-, c-, D-, E- and T-optimality criteria, several numerical algorithms have been developed for finding the optimal designs, and they include a
multiplicative algorithm in Bose and Mukerjee (2015),  
a general and efficient algorithm in Yang et al. (2013),
a cocktail algorithm in Yu (2011),
algorithms based on semi-definite programming in Atashgah and Seifi (2007), 
Papp (2012), Ye et al. (2017),
and Wong et al. (2019), 
and algorithms based on semi-infinite programming in Duarte et al. (2015).

CVX (Grant and Boyd, 2013) is a user-friendly package for MATLAB which uses solvers such as SeDuMi (Sturm, 1999) to solve a large class of constrained convex optimization problems. Many papers have applied CVX to find various optimal designs; see e.g. Atashgah and Seifi (2007), Papp (2012), and Wong et al. (2019). While the R-optimality problem we consider is a  convex optimization problem (Theorem \ref{Th1}), it cannot be solved by CVX. 

Lu and Pong (2013) applied an interior point method to find A- and D-optimal designs. Here, we propose using an interior point method to compute R-optimal designs.  The R-optimal design problem on $S_N$ is defined as follows: 
\begin{eqnarray} 
\min_{\bf w}~ \{  \phi({\bf w}) \} ~~~\mbox{subject to} ~~~w_j \geq 0, ~\sum_{j=1}^N w_j=1,  \label{Prob1}
\end{eqnarray}
where $\phi({\bf w})$ is defined in (\ref{logloss}). This is a convex optimization problem with equality and inequality constraints.  
To deal with the inequality constraints, we use a log-barrier function,
\begin{eqnarray}
h({\bf w},t) =-\frac{1}{t} \sum_{j=1}^N \log{(w_j)}, ~~\mbox{with}~~t>0,
\label{gBarrier}
\end{eqnarray}
to form another convex optimization problem,
\begin{eqnarray}
\min_{\bf w} ~\{\phi_1({\bf w},t)\} ~~~ \mbox{subject to} ~~~\sum_{j=1}^N w_j=1, \label{Prob2} 
\end{eqnarray}
where $\phi_1({\bf w},t)=\phi({\bf w}) + h({\bf w},t)$.  For fixed $t$, the solution to problem
(\ref{Prob2}) must satisfy the constraints in problem (\ref{Prob1}) due to the 
log-barrier function $h({\bf w},t)$. Let ${\bf w}^{(t)}$ be the solution to problem (\ref{Prob2}).  The solution to problem (\ref{Prob1}) is exactly equal to the limit of  ${\bf w}^{(t_k)}$, as $t_k \to \infty$  (Boyd and Vandenberghe 2004, p.564-566). Thus, to solve problem (\ref{Prob1}), we can simply solve problem (\ref{Prob2}) for an increasing sequence of $t$, say, $t_1 < t_2 < \ldots$ until convergence. We say that the algorithm has converged when \eqref{eqv2} is satisfied. 

To solve problem (\ref{Prob2}), we remove the equality constraint by replacing $w_N$ by $1-\sum_{j=1}^{N-1}w_j$ in ${\bf w}$. Let $\tilde{\bf w}=(w_1, \ldots, w_{N-1}, 1-\sum_{j=1}^{N-1}w_j)$. Then, for fixed $t$,  $\phi_1(\tilde{\bf w},t)$ is a convex function of $w_1, \ldots, w_{N-1}$. Let ${\bf g}(\tilde{\bf w},t) $ be the gradient vector of $\phi_1(\tilde{\bf w},t)$ with respect to $w_1, \ldots, w_{N-1}$.  The formula of ${\bf g}(\tilde{\bf w},t) $ is given in the Appendix.  We can now apply the BFGS (Broyden--Fletcher--Goldfarb--Shanno) Algorithm for solving unconstrained optimization problem (\ref{Prob2}); see Antoniou and  Lu (2007, p191-192) for a detailed description of the BFGS algorithm. The BFGS algorithm only requires the gradient vector of $\phi_1(\tilde{\bf w},t)$. Furthermore, the BFGS algorithm is more computationally efficient than Newton's method when $N$ is large. 
Unlike Newton's method, the BFGS algorithm does not require computation of the $(N-1) \times (N-1)$ Hessian matrix of $\phi_1(\tilde{\bf w},t)$; 
instead, the BFGS algorithm uses a sequence of approximations to the inverse of the Hessian matrix.  
In the BFGS algorithm we also make an adjustment to search for optimal step sizes so that all the elements of $\tilde{\bf w}$ are positive. The details of the algorithm for computing R-optimal designs are given in Algorithm 1. 

\begin{algorithm}
\caption{Interior point method for computing R-optimal designs}
\begin{enumerate} 
\item  Define an initial weight vector ${\bf w}^{(t_0)} = \frac{1}{N} {\bf 1}_N$, and let $t_1 > 0$ (say, $t_1 = 2$). Fix a small positive $\delta$, say, $\delta=10^{-8}$. Let  $\lambda > 1$ (say, $\lambda = 2$). 
\item For $k = 1, 2, \ldots$
\begin{enumerate} 
\item Solve problem (\ref{Prob2}) with $t=t_k$ using BFGS, initializing the algorithm with ${\bf w}^{(t_{k - 1})}$, and denoting the solution as ${\bf w}^{(t_k)}$.
\item Compute $d({\bf w}^{(t_k)},j)$ defined in (\ref{eqv1}) for $j=1, \ldots, N$. 
\item  If $\max_{j} d({\bf w}^{(t_k)},j) \le \delta$, then stop and ${\bf w}^{(t_k)}$ is an R-optimal design. Otherwise, let $t_{k+1}= \lambda t_k$. 
\end{enumerate} 
\end{enumerate} 
\end{algorithm}	

The parameter $\lambda$ in Algorithm 1 is used to generate the increasing sequence $t_1, t_2, \ldots$.
This parameter may need to be adjusted to compute optimal designs for other models. 
See Boyd and Vandenberghe (2004, p570) for detailed discussion on the choice of $\lambda$.
We found that $\lambda=2$ worked well for all
the examples in Section 4.
 
In Section 4, we compare Algorithm 1 to a multiplicative algorithm for problem (\ref{Prob1}), and find that Algorithm 1 works better for complicated models and
large $N$. 

%
%
%
%
%


\section{Applications}

We are able to find R-optimal designs easily on discrete design spaces using Algorithm 1.
We use three examples to present results for R-optimal designs,  compare Algorithm 1 to a
multiplicative algorithm, and address a few numerical issues. Example 1 is discussed in Section 2, which is a 3-response linear regression model with 2 design variables
and $q=15$.
In Example 2 we have a 3-response linear regression model with five design variables (including two
categorical variables) and $q=28$, where $N$ can be huge.
Example 3 is for a 2-response nonlinear regression model.
All the compuation is done on a PC equipped with 
Intel Core i7-8700 Six Core 4.6 GHz CPU 16 GB 2666 MHz DDR4.

\noindent{\bf Example 1} [Continued] We compute R-optimal designs on $S_{N,1}$ and 
 $S_{N,2}$ with   two  matrices for ${\bf V}_0$ given by
${\bf V}_{0,1}=\left( \begin{array}{ccc}	 
	4 & 3 & 4 \\
	3 & 9 & 6 \\
	4 & 6 & 16 
	\end{array} \right)$ and 
${\bf V}_{0,2}=\left( \begin{array}{ccc}	 
	4 & 1.8 & 2.5 \\
	1.8 & 9 & 10.6 \\
	2.5 & 10.6 & 56 
\end{array} \right).$
When ${\bf V}_0$ is given by ${\bf V}_{0,1}$, the response variables $y_1, y_2$ and $y_3$ have the same pairwise correlations (0.5). When ${\bf V}_0$ is given by ${\bf V}_{0,2}$, $y_1, y_2$ and $y_3$ have different pairwise correlations. Representative R-optimal designs and computation times are given in Table \ref{Table1}. The computation times may vary over different runs and ${\bf V}_0$, but they are usually within 2 seconds of the reported times.  When we increase $N$ from $15^2$ to $21^2$, the computation times increase to about 95 seconds.

[Table 1 near here]

From Table \ref{Table1}, observe that the R-optimal designs all have 9 support points. Furthermore, the R-optimal designs on  $S_{N,2}$ have the reflection symmetry property 
 while those on  $S_{N,1}$ have the line symmetry property, as
discussed in Section 2. We use $\delta = 10^{-8}$ in Step 1 of Algorithm 1. A representative 3-D plot of $d({\bf w}^*,j)$ on $S_{N,2}$ is given in Figure 1, which also shows the 9 support points clearly.
For a multiplicative algorithm, it takes about 27 and 86 seconds for $N=15^2$ and $21^2$, respectively. 
Thus, both algorithms are very computationally efficient, and the multiplicative algorithm works well for simple models and small $N$.

[Figure 1 near here] 

We have computed R-optimal designs for various ${\bf V}_0$ on a fixed  design space $S_N$ and noticed that the support points do not change, but
the weights change slightly. 
\hfill{$\Box$}

Algorithm 1 can find R-optimal designs for models including both continuous and  categorical variables, and all the properties derived in Section 2 also hold for these R-optimal designs.  Example 2 is used to illustrate those points and to show that  Algorithm 1 works well for many design variables and large  $N$.

\noindent{\bf Example 2} Consider a 3-response linear model with 5 design variables,
\begin{eqnarray*}
&&{\bf f}_1({\bf x})=(1, x_1, x_2, x_3, x_4, x_5, x_1x_4, x_1x_5,  x_2x_4, x_2x_5, x_3x_4, x_3x_5)^\top, \\
&&{\bf f}_2({\bf x})=(1, x_1, x_2, x_3, x_4, x_5, x_1x_4, x_1x_5, x_1x_2, x_1x_3 )^\top, \\
&&{\bf f}_3({\bf x})=(1, x_1, x_2, x_3, x_4, x_5)^\top, 	
\end{eqnarray*}	
where $x_1 \in [-1, 1], x_2 \in [-1, 1], x_3 \in [-1, 1], x_4=0,1, x_5=0, 1$.  Variables $x_1, x_2$ and $x_3$ are continuous, while $x_4$ and $x_5$ are categorical variables coded as $0$ (for baseline) and $1$.  The total number of regression parameters is $q=28$.
We can use $N_1, N_2$ and $N_3$ grid points to discretize interval $[-1,1]$ for $x_1, x_2$ and $x_3$ respectively, so design space
$S_N$ has $N=4N_1N_2N_3$ points.  
We consider $N_1=N_2=N_3=N_*$ with various values of $N_*$: 6, 8, 10 and 16,
and the corresponding values of $N$ are, respectively, 864, 2048, 4000 and 16384.
Using Theorem \ref{Th4}, we can focus on finding an R-optimal 
design having reflection symmetry with respect to $x_1, x_2$ and $x_3$.  Thus, we can reduce the number of unknown weights
in problem (\ref{Prob2}) to $N/8$, which significantly speeds up computation time.

The R-optimal design is the same for all the values of $N_*$, and it has 32 support points. We only present 4 support points in Table \ref{table2}, since the remaining 28 support points can be easily obtained  by the reflection symmetry with respect to $x_1, x_2$ and $x_3$.  
\hfill{$\Box$}

[Table 2 near here]

Although we focus on multi-response linear regression models in
Sections 2 and 3, all the results can be extended to multi-response nonlinear regression models easily and the algorithm can be applied to find R-optimal designs.  For a nonlinear response, say
$E(y)=g({\bf x}, {\bm \beta})$, we define ${\bf f}({\bf x}, {\bm \beta})$ to be
the first derivative of  $g({\bf x}, {\bm \beta})$ 
with respect to ${\bm \beta}$. Suppose ${\bm \beta}_*$ is the true parameter value.  We then use ${\bf f}({\bf x}, {\bm \beta}_*)$ in matrix $Z({\bf x})$ to calculate $Cov(\hat{\bm \beta}) $
in (\ref{Cove2}).  It is clear that the R-optimal designs depend on ${\bm \beta}_*$ and they are  called 
locally R-optimal designs.  We illustrate these ideas in Example 3.

\noindent{\bf Example 3} Consider a 2-response nonlinear regression model given by
\begin{eqnarray*} y_1=\frac{\beta_{11} x}{x+\beta_{12}} + \epsilon_1, \quad 
y_2=\frac{\beta_{21} x}{x+\beta_{22}} + \epsilon_2,
\end{eqnarray*}
where design variable $x \in [0, b]$, regression parameter vector ${\bm \beta}_1=(\beta_{11}, \beta_{11})^\top$,
${\bm \beta}_2=(\beta_{21}, \beta_{22})^\top$, and the errors have mean zero and $Cov((\epsilon_1,\epsilon_2)^\top)={\bf V}_0$.
This is a bivariate Emax model which is often used to investigate the efficacy and side-effects of a drug; see e.g.   Magnusdottir (2013) for an application of the model. Variable
$x$ denotes the dose level of a drug.  Locally R-optimal designs are computed for various true values of
${\bm \beta}_1, {\bm \beta}_2$ and ${\bf R}_0$, where ${\bf R}_0$ is given by
${\bf R}_0=\left(  \begin{array}{cc}
1 & \rho  	 \\
\rho   & 1
\end{array}
\right).$
%
Since the response functions are linear in parameters $\beta_{11}$ and $\beta_{21}$, the R-optimal designs do not depend on their true 
values, and so we set them to  be 1.

[Table 3 near here]

Table \ref{table3} gives representative results from Algorithm 1, where we define $S_N$ to be $N$ equally spaced grid points in $[0, b]$.  These results indicate that 
(i) the R-optimal designs depend on $\rho$ in ${\bf R}_0$ through the absolute value of $\rho$,
(ii) the boundary point $b$ is always a support point in the R-optimal designs,
(iii) the number of support points in the R-optimal designs is either 2 or 3,
(iv) Algorithm 1 converges quickly to the  R-optimal designs, where $\delta=10^{-8}$ is used in Step 1 of Algorithm 1.
The multiplicative algorithm is very slow for this nonlinear model, and it does not converge for several cases
listed in Table \ref{table3}.
\hfill{$\Box$}

For some applications, matrix ${\bf I}({\bf w})$ can be ill-conditioned. This can be problematic, as we need to compute the inverse of ${\bf I}({\bf w})$ in Algorithm 1. 
However, we can scale the design space $S_N$ or scale the covariance matrix ${\bf V}_{0}$ so that
it is easier to find the inverse of ${\bf I}({\bf w})$.
Similar ideas are discussed in Wong and Zhou (2019) for CVX based algorithms.

\section{Conclusion}

In this paper, we studied R-optimal designs for both linear and nonlinear multi-response models and derived various theoretical results. Although we have used discrete design spaces to formulate the design problem and obtain the theoretical results, the theoretical results also hold for compact design spaces, and the proofs are similar. 

Throughout this paper, we have considered optimal designs based on the GLSE $\hat{\bm \beta}$ given in \eqref{glse}, as it is the BLUE for $\bm \beta$. However, the GLSE depends on the error covariance matrix ${\bf V}_0$, which is often unknown in practice. One option is to use the feasible GLSE, which replaces ${\bf V}_0$ in \eqref{glse} with an estimate ${\hat{\bf V}}_0$; when $\hat {\bf V}_0$ is a consistent estimator of ${\bf V}_0$, the asymptotic covariance matrix of the feasible GLSE converges to the covariance matrix of the GLSE. 
Alternatively, we may instead construct R-optimal designs based the least squares estimator (LSE) for ${\bm \beta}$. However, the loss function $\phi({\bf w})$ based on the LSE is not a convex function of ${\bf w}$, and thus, computing the R-optimal designs based on the LSE is more challenging. 

When ${\bf V}_0$ or ${\bf R}_0$ are misspecified, the R-optimal designs may be very inefficient. Thus, it is of interest to consider robust R-optimal designs. A minimax approach is a possible way to deal with misspecification of ${\bf V}_0$ or ${\bf R}_0$. This approach is challenging, as the resulting design problem has a non-convex and possibly non-smooth objective function.



\section*{Acknowledgements}

This research work was partially supported by Discovery Grants from the Natural
Sciences and Engineering Research Council of Canada.

\section*{References} 
{\small

\begin{description}

\item Antoniou, A. and Lu, Wu-Sheng (2007).
{\it Practical Optimization Algorithms and Engineer Applications}.
Springer, New York.	

\item Atashgah, A.B. and Seifi, A. (2007).
Application of semi-definite programming to the design of multi-response experiments.
{\it IIE Transactions}, 39,  763-769.

\item Atashgah, A.B. and Seifi, A. (2009).
Optimal design of multi-response experiments using semi-definite programming.
{\it Optimization and Engineering}, 10, 75-90.

\item Bose, M. and Mukerjee, R. (2015).
Optimal design measures under asymmetric errors, with
application to binary design points.
{\it Journal of Statistical Planning and Inference}, 159, 28-36.

\item Boyd, S. and  Vandenberghe, L. (2004).
{\it Convex Optimization}.
Cambridge University Press, New York.

\item Dette, H. (1997).
Designing experiments with respect to ‘standardized’ optimality criteria.
{\it Journal of Royal Statistical Society B}, 59, 97-110.

\item Duarte, B.P.M., Wong, W.K. and Atkinson, A.C. (2015).
A semi-infinite programming based algorithm for determining T-optimum designs for model discrimination.
{\it Journal of Multivariate Analysis}, 135, 11-24.


\item Grant, M.C. and  Boyd, S.P. (2013).
{\it The CVX Users' Guide}.
Release 2.0 (beta),
CVX Research, Inc. (http://cvxr.com/cvx/doc/CVX.pdf, October 14, 2013.)

\item He, L. and Yue, R.X. (2018).
R-optimal designs for trigonometric regression models.
{\it Statistical Papers}, published online https://doi.org/10.1007/s00362-018-1017-x.

\item He, L. and Yue, R.X. (2019).
R-optimality criterion for regression models with asymmetric errors.
{\it Journal of Statistical Planning and Inference}, 199, 318-326.

\item Kiefer, J. (1974).  General equivalence theorem for optimum designs (approximate theory).
{\it The Annals of Statistics}, 2, 849-879.


\item Liu, X. and Yue, R.X. (2013).
A note on R-optimal designs for multiresponse models.
{\it Metrika}, 76, 483-493.

\item Liu, X., Yue, R.X. and Chatterjee, K. (2014).
A note on R-optimal designs for multi-factor models.
{\it Journal of Statistical Planning and Inference}, 146, 139-144.

\item Liu, X., Yue, R.X. and Chatterjee, K. (2014).
R-optimal designs in random coefficient regression models.
{\it Statistics and Probability Letters}, 88, 127-132.

\item Liu, X., Yue, R.X., Xu, J. and Chatterjee, K. (2016).
Algorithmic construction of R-optimal designs for second-order response surface models.
{\it Journal of Statistical Planning and Inference}, 178, 61-69.

\item Lu, Z.S. and Pong, T.K. (2013).
Computing optimal experimental designs via interior point method.
{\it SIAM Journal on  Matrix Analysis and  Applications}, 34, 1556-1580.

\item Magnusdottir, B.T. (2013). c-Optimal designs for the bivariate Emax model. In
{\it mODa 10-Advances in Model-Oriented Design and Analysis}, edited by Uci\'{n}ski, D.,
Atkinson, A.C., and Patan, M., page 153-161.  Springer, Switzerland.


\item Papp, D. (2012).
Optimal designs for rational function regression.  
{\it Journal of the American Statistical Association}, 107, 400-411.

\item Sturm, J. (1999). Using SeDuMi 1.02, a MATLAB toolbox for optimization over symmetric cones. {\it Optimization Methods Software}, 11, 625-653.

\item Wong, W.K., Yin, Y. and Zhou, J. (2019).
Optimal designs for multi-response nonlinear regression models with several factors via semi-definite programming.
{\it Journal of Computational and Graphical Statistics}, 28, 61-73.

\item Wong, W.K. and Zhou, J. (2019).
CVX based algorithms for constructing various optimal regression designs.
{\it Canadian Journal of Statistics}, to appear.

\item Yang, M., Biedermann, S. and Tang, E. (2013).
On optimal designs for nonlinear models:
a general and efficient algorithm.
{\it Journal of the American Statistical Association}, 108, 1411-1420.

\item Ye, J.J., Zhou, J. and Zhou, W. (2017).
Computing A-optimal and E-optimal designs for regression models via semidefinite programming.
{\it Communications in Statistics - Simulation and Computation}, 46, 2011-2024.

\item Yin, Y. and Zhou, J. (2017).
Optimal designs for regression models using the second-order least squares estimator.
{\it Statistica Sinica}, 27, 1841-1856.

\item Yu, Y. (2011).
D-optimal designs via a cocktail algorithm.
{\it Statistics and Computing}, 21, 475-481.

\end{description}             

}


\section*{Appendix: Proofs and derivations}

\noindent{\bf Proof of Theorem 1:}
Since ${\bf I}({\bf w}_1)$ and ${\bf I}({\bf w}_2)$ are non-singular, it is clear that
$ {\bf I}({\bf w}_{\alpha})=(1-\alpha) {\bf I}({\bf w}_1) + \alpha {\bf I}({\bf w}_2) $ is also 
non-singular for all $\alpha \in [0,1]$ and ${\bf A}({\bf w}_{\alpha})$ exists.
By (\ref{logloss}), 
\begin{eqnarray}
\frac{d \phi({\bf w}_{\alpha})}{d \alpha} &=&
\sum_{r=1}^q \frac{-1}{{\bf e}_r^\top {\bf A}({\bf w}_{\alpha}){\bf e}_r }
{\bf e}_r^\top {\bf A}({\bf w}_{\alpha}) \left[ {\bf I}({\bf w}_2) - {\bf I}({\bf w}_1) \right]
{\bf A}({\bf w}_{\alpha}) {\bf e}_r, 
\label{firstdiv} \\ 
\frac{d^2 \phi({\bf w}_{\alpha})}{d \alpha^2} &=& \sum_{r=1}^q 
\frac{1}{\left({\bf a}_r^\top{\bf a}_r\right)^2}
\left( 2 {\bf a}_r^\top{\bf a}_r \cdot {\bf b}_r^\top{\bf b}_r - 
\left({\bf a}_r^\top{\bf b}_r\right)^2 \right), \nonumber
\end{eqnarray}
where ${\bf a}_r = \left( {\bf A}({\bf w}_{\alpha}) \right)^{1/2} {\bf e}_r$
and ${\bf b}_r = \left( {\bf A}({\bf w}_{\alpha}) \right)^{1/2} 
\left[ {\bf I}({\bf w}_2) - {\bf I}({\bf w}_1) \right] {\bf A}({\bf w}_{\alpha}) {\bf e}_r$.
By the Cauchy-Schwarz inequality, $\frac{d^2 \phi({\bf w}_{\alpha})}{d \alpha^2} \ge 0$ for all $\alpha \in [0,1]$, so
$\phi({\bf w}_{\alpha})$ is a convex function of $\alpha$.
\hfill{$\Box$}

\vspace{3mm} 

\noindent{\bf Proof of Theorem 2:} Suppose ${\bf w}^*$ is an R-optimal design and ${\bf w}$ is another design.  
Let ${\bf w}^*_{\alpha}=(1-\alpha) {\bf w}^* + \alpha {\bf w}$.
Then, $ \frac{d \phi({\bf w}^*_{\alpha})}{d \alpha}  \mid_{\alpha=0} \ge 0$, for any {\bf w}.
From (\ref{firstdiv}) it is easy to obtain 
$d({\bf w}^*, j) \le 0$ for all $j=1, \ldots, N$, and the equality holds at the support points of 
$\xi({\bf w}^*)$.
\hfill{$\Box$}
\bigskip

\noindent{\bf Proof of Theorem 3:}  
On $S_N^T$, let 
$${\bf I}_T({\bf w}) =\sum_{j=1}^N w_j \left(Z(T{\bf u}_j) \right)^\top {\bf V}_0^{-1} Z(T{\bf u}_j)
~~\mbox{and} ~~{\bf A}_T({\bf w}) = \left({\bf I}_T({\bf w})\right)^{-1}.$$
We minimize $\phi_T({\bf w}) =\sum_{r=1}^q \log({\bf e}_r^\top {\bf A}_T({\bf w}){\bf e}_r)$ to get an R-optimal design on $S_N^T$. 

Using the assumption in Theorem \ref{Th3}, we can easily show that
${\bf A}_T({\bf w})=Q^{-1} {\bf A}({\bf w}) Q^{-1}$ and
$\phi_T({\bf w}) = \sum_{r=1}^q \log \left(q_{rr}^2\right) + \phi({\bf w})$,
where $q_{rr}$ are the diagonal elements of $Q^{-1}$.
Thus, if ${\bf w}^*$ minimizes $\phi({\bf w})$, then it also minimizes $\phi_T({\bf w})$.
This implies that the R-optimal design for the model is scale invariant under 
transformation $T$.
\hfill{$\Box$}
\bigskip

\noindent{\bf Proof of Theorem 4:} Let $q_{rr}$ be the diagonal elements of $Q^{-1}$. We can write the loss function on $S_N^{T_l}$ as
$\phi_{T_l}({\bf w}) = \sum_{r=1}^q  log(q_{rr}^2) + \phi({\bf w}).$ Since $q_{rr}=\pm 1$ for all $r$, $\phi_{T_l}({\bf w}) = \phi({\bf w})$ for 
all ${\bf w}$.  Furthermore, since $S_N$ has a reflection symmetry with respect to $x_l$,  $S_N^{T_l} = S_N$.
Finally, since  $\phi_{T_l}({\bf w}) = \phi({\bf w})$,  if $\xi({\bf w}^*)$ is an R-optimal design on $S_N$,
then $\tilde{\xi}({\bf w}^*)$ is also an R-optimal design on $S_N$.
If $\xi({\bf w}^*)$ and $\tilde{\xi}({\bf w}^*)$ are the same, then 
 $\xi({\bf w}^*)$ has the reflection symmetry  with respect to $x_l$ 
 and the result is proved.
 Otherwise, we can take a convex combination of $\xi({\bf w}^*)$ and $\tilde{\xi}({\bf w}^*)$,
 $0.5 \xi({\bf w}^*) + 0.5 \tilde{\xi}({\bf w}^*)$,
 which has the reflection symmetry  with respect to $x_l$ and is an R-optimal design by convexity
 of $\phi({\bf w}) $ (Theorem \ref{Th1}).
\hfill{$\Box$}
\bigskip

\noindent{\bf Proof of Theorem 6:} Substituting $ {\bf V}_0= {\bm \Sigma}_0 {\bf R}_0 {\bm \Sigma}_0$ into \eqref{matrixA}, we get 
\begin{eqnarray*}
{\bf I}({\bf w})  &=&\sum_{j=1}^N w_j U_j^\top {\bm \Sigma}_0^{-1} {\bf R}_0^{-1} {\bm \Sigma}_0^{-1} U_j \\
                 &=&\sum_{j=1}^N w_j {\bf Q}_1 U_j^\top  {\bf R}_0^{-1}  U_j {\bf Q}_1,  
             ~~\mbox{where} ~{\bf Q}_1=\frac{1}{\sigma_1}{\bf I}_{q_1} \oplus \cdots \oplus \frac{1}{\sigma_m}{\bf I}_{q_m}, \\
                 &=& {\bf Q}_1 \left( \sum_{j=1}^N w_j  U_j^\top  {\bf R}_0^{-1}  U_j  \right) {\bf Q}_1, 
\end{eqnarray*}
It follows that ${\bf A}({\bf w}) = {\bf Q}_1 ^{-1}  \left( \sum_{j=1}^N w_j  U_j^\top  {\bf R}_0^{-1}  U_j  \right)^{-1} {\bf Q}_1^{-1}.$
By (\ref{logloss}) and the proof of Theorem 3, it is clear that the R-optimal design for model (\ref{model10}) depends on
$ {\bf V}_0$  only through $ {\bf R}_0$.
\hfill{$\Box$}


\bigskip

\noindent{\bf Formula for ${\bf g}(\tilde{\bf w},t)$:}  Notice that
${\bf g}(\tilde{\bf w},t) = \nabla 	\phi_1(\tilde{\bf w},t) 
	                         = \nabla 	\phi(\tilde{\bf w}) + \nabla 	h(\tilde{\bf w},t).$ \\
For $j=1, \ldots, N$, let ${\bf B}_j=U_j^\top {\bf V}_0^{-1} U_j$.  
By (\ref{matrixA}) (\ref{logloss}), and (\ref{gBarrier}),  for $j=1, \ldots, N-1$,
\begin{eqnarray*}
\frac{\partial 	\phi(\tilde{\bf w})}{\partial w_j} = \sum_{r=1}^q 
\frac{{\bf e}_r^\top {\bf A}(\tilde{\bf w}) \left({\bf B}_N-{\bf B}_j \right) {\bf A}(\tilde{\bf w}) {\bf e}_r}
{{\bf e}_r^\top {\bf A}(\tilde{\bf w}) {\bf e}_r}, \quad \text{and} \quad  \frac{\partial 	h(\tilde{\bf w},t)}{\partial w_j} = \frac{1}{t} \left(
 \frac{1}{1-\sum_{i=1}^{N-1}w_i} 
 -\frac{1}{w_j}  \right).
\end{eqnarray*}	
\hfill{$\Box$}

\newpage


\begin{small}
\begin{table}[H]
	\caption{R-optimal designs and computation times from Algorithm 1 for Example 1}	
	\begin{center}
		\begin{tabular}{lrrrr}
			Design space & support point & weight for ${\bf V}_{0,1}$ & weight for ${\bf V}_{0,2}$& time (sec.) \\ \hline
			$S_{N,1}$ &(0.0000, 0.0000) &   0.2500 & 0.2530 & 23.2188 \\ 
			$N_1N_2=15^2$ & (0.0000,    0.5000)   &    0.1242  & 0.1235 & \\ 
			& (0.0000,    1.0000)   &     0.0864 & 0.0856 & \\
			& (0.5000,    0.0000)   &     0.1242 & 0.1235 & \\
			& (0.5000,    0.5000)   &     0.1100 & 0.1108 & \\
			& (0.5000,    1.0000)   &     0.0678 & 0.0680 & \\
			& (1.0000,    0.0000)   &     0.0864 & 0.0856 & \\
			& (1.0000,    0.5000)   &     0.0678 & 0.0680 & \\
			& (1.0000,    1.0000)   &     0.0832 & 0.0820 & \\ \hline
			$S_{N,2}$ &  ($-1.0000,   -5.0000$) &    0.1305 & 0.1297 & 23.2969  \\ 
			$N_1N_2=15^2$ & ($-1.0000,   0.0000$) &    0.0822 & 0.0822 & \\
			& ($-1.0000,    5.0000$)&    0.1305 &  0.1297 & \\
			& ($0.0000,   -5.0000$) &    0.0822 & 0.0822 & \\
			& ($0.0000,    0.0000$) &    0.1492 & 0.1524 & \\
			& ($0.0000,    5.0000$) &    0.0822 & 0.0822 & \\
			& ($1.0000,   -5.0000$) &    0.1305 & 0.1297 & \\
			& ($1.0000,    0.0000$) &    0.0822 & 0.0822 & \\
			& ($1.0000,    5.0000$) &    0.1305 & 0.1297 & \\ \hline
		\end{tabular}
	\end{center}
	\label{Table1}
\end{table}
\end{small}

\begin{table}[H]
\caption{Support points and weights in the R-optimal design for Example 2 with ${\bf V}_{0,1}$.}
\begin{center}
\begin{tabular}{rrrrrr} \hline
~~~~~$x_1$   & 	~~~~~$x_2$   & ~~~~~$x_3$  & ~~~$x_4$   & ~~~$x_5$  & weight   \\  \hline
$+1$  & $+1$  &  $+1$  & 0  & 0  &  0.0511\\
$+1$  & $+1$  &  $+1$  & 0  & 1  &  0.0263\\
$+1$  & $+1$  &  $+1$  & 1  & 0  &  0.0263\\
$+1$  & $+1$  &  $+1$  & 1  & 1  &  0.0213\\  \hline
\multicolumn{5}{l}{Computation time (sec.) for $N=16384$:} & 1801.3157 \\ \hline
\end{tabular}
\end{center}
\label{table2}	
\end{table}

\begin{table}[H]
	\caption{Locally R-optimal designs for bivariate Emax model.}
	\begin{center}
	\begin{tabular}{lcrrr}  
Design space & true parameter & support point & weight & time (sec.) \\ \hline
$b=100$	 & 	${\bm \beta}_1=(1,1)^\top $ &      1.0000 &  0.2532 & 4.3901 \\
$N=101$      &  ${\bm \beta}_2=(1,5)^\top $ &  4.0000     &  0.2138  &    \\
&  $\rho= \pm 0.5$  &  100.0000      &  0.5330 &    \\ \hline
$b=100$	 & 	${\bm \beta}_1=(1,1)^\top $ &  1.0000     & 0.2635  &  15.2377 \\
$N=201$      &  ${\bm \beta}_2=(1,5)^\top $ &   4.5000    &  0.2075  &    \\
              &  $\rho= \pm 0.5$  &  100.0000     & 0.5290  &    \\ \hline   
$b=100$	 & 	${\bm \beta}_1=(1,1)^\top $ &  1.0000     & 0.2617  & 117.1706 \\
$N=501$      &  ${\bm \beta}_2=(1,5)^\top $ &   4.4000    &  0.2086 &    \\
&  $ \rho=\pm 0.5$  &    100.0000   & 0.5297  &    \\ \hline   
$b=150$	 & 	${\bm \beta}_1=(1,1)^\top $ &  1.5000     & 0.3187  &  30.5318 \\
$N=301$      &  ${\bm \beta}_2=(1,5)^\top $ &   3.5000    & 0.1218  &    \\
&  $\rho= \pm 0.3$  &   150.0000    & 0.5595  &    \\ \hline    
$b=150$	 & 	${\bm \beta}_1=(1,3)^\top $ &    2.5000   & 0.2731  & 31.7188  \\
$N=301$      &  ${\bm \beta}_2=(1,10)^\top $ &   9.5000    & 0.2020  &    \\
&  $ \rho= \pm 0.7$  &   150.0000    & 0.5249  &    \\ \hline    
$b=150$	 & 	${\bm \beta}_1=(1,3)^\top $ &    4.2000   & 0.4492  & 119.3701  \\
$N=501$      &  ${\bm \beta}_2=(1,10)^\top $ &   150.0000    & 0.5508  &    \\
&  $ \rho= \pm 0.1$  &        &    &    \\ \hline                                  
\end{tabular} 
\end{center}
\label{table3}         
\end{table}

\begin{figure}[H]
	\centering
	\includegraphics[width=3.8in]{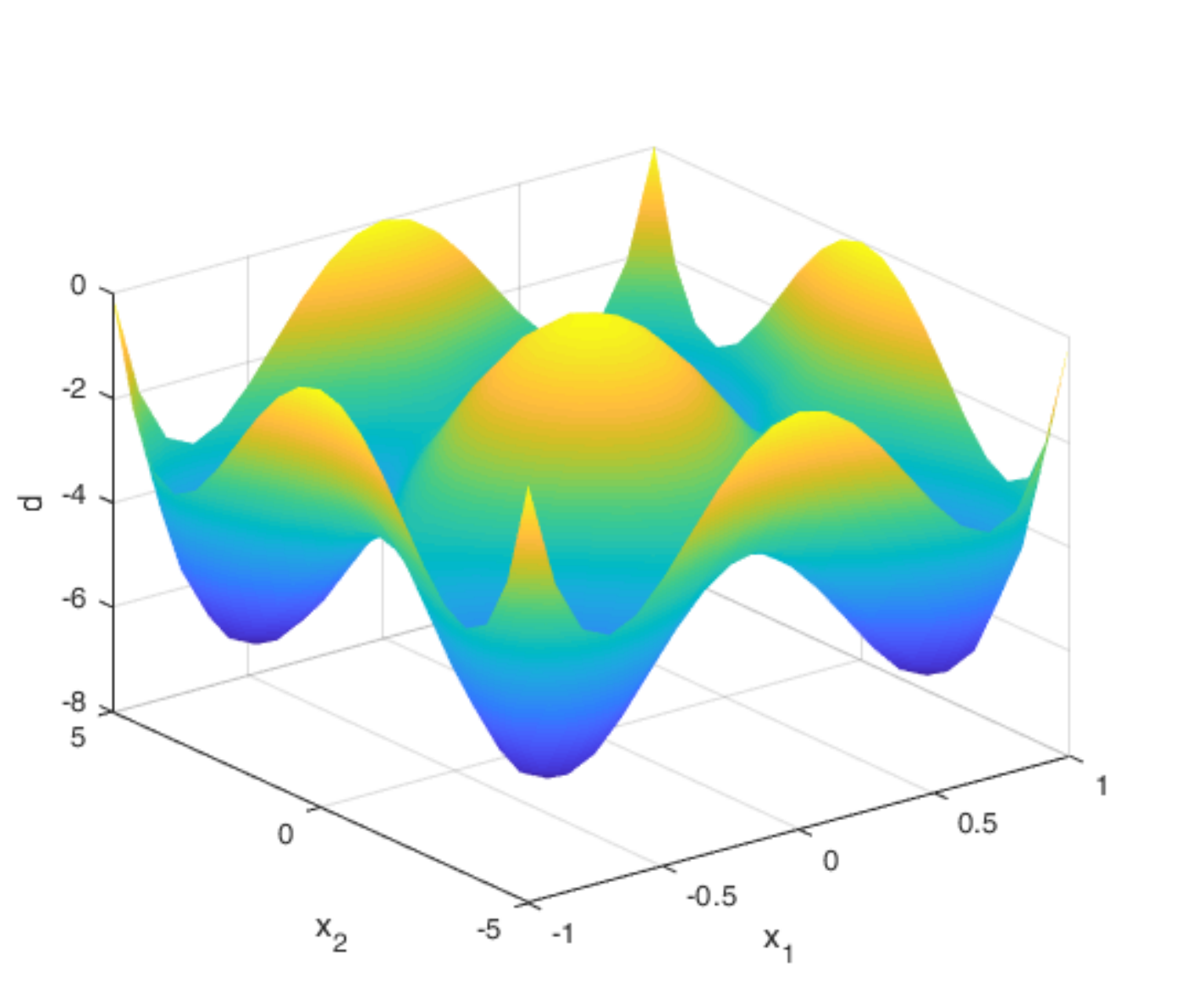}
	\caption{\vspace{-2cm}A plot of $d({\bf w}^*,j)$ versus $(x_1,x_2) $ ($\in S_{N,2}$)  with ${\bf V}_{0,2}$ for Example 1.}
\end{figure}



\end{document}